\documentclass[12pt]{article}
\pdfoutput=1

\usepackage{amsmath}
\usepackage{amssymb}
\usepackage{epsfig}
\usepackage{graphicx}
\usepackage{cite}
\usepackage{multirow}
\usepackage{longtable}
\usepackage{lscape}
\usepackage{bbm}
\usepackage{dcolumn}
\usepackage{rotating}

\allowdisplaybreaks[4] 

\newcommand{\capdef}{}
\newcommand{\mycaption}[2][\capdef]{\renewcommand{\capdef}{#2}%
        \caption[#1]{{\footnotesize #2}}}
\makeatletter
\renewcommand{\fnum@table}{\textbf{\tablename~\thetable}}
\renewcommand{\fnum@figure}{\textbf{\figurename~\thefigure}}
\makeatother

\newcounter{myenumi}

\renewcommand{\themyenumi}{\roman{myenumi}}
{\end{list}}

\newlength{\myem}
\newcounter{mysubequation}[equation]

\makeatletter
\renewcommand{\section}{\@startsection{section}{1}{0em}{-\baselineskip}%
{\baselineskip}{\normalfont\large\bfseries}}
\renewcommand{\subsection}%
{\@startsection{subsection}{2}{0em}{-0.7\baselineskip}%
{0.7\baselineskip}{\normalfont\bfseries}}
\makeatother

\newcommand{\bi}{\begin{itemize}}
\newcommand{\ei}{\end{itemize}}

\newcommand{\be}{\begin{equation}}
\newcommand{\ee}{\end{equation}}
\newcommand{\bea}{\begin{eqnarray}}
\newcommand{\eea}{\end{eqnarray}}

\newcommand{\ie}{{\it i.e.}}

\newcommand{\eg}{{\it e.g.}}

\newcommand{\cf}{{\it cf.}}

\newcommand{\eq}{Eq.}

\newcommand{\fig}{Fig.}

\newcommand{\Ref}{Ref.}
\newcommand{\Refs}{Refs.}
\newcommand{\Sec}{Sec.}

\newcommand{\App}{the Appendix}

\newcommand{\equ}[1]{\eq~(\ref{equ:#1})}
\newcommand{\figu}[1]{\fig~\ref{fig:#1}}

\begin{document}


\begin{titlepage}

\renewcommand{\thefootnote}{\alph{footnote}}

\vspace*{-3.cm}
\begin{flushright}

\end{flushright}


\renewcommand{\thefootnote}{\fnsymbol{footnote}}
\setcounter{footnote}{-1}

{\begin{center}
{\large\bf
 Interpretation of neutrino flux limits from neutrino telescopes \\ on the Hillas plot
}
\end{center}}

\renewcommand{\thefootnote}{\alph{footnote}}

\vspace*{.8cm}
\vspace*{.3cm}
{\begin{center} {\large{\sc
                Walter Winter\footnote[1]{\makebox[1.cm]{Email:}
                winter@physik.uni-wuerzburg.de}
                }}
\end{center}}
\vspace*{0cm}
{\it
\begin{center}

\footnotemark[1]%
       Institut f{\"u}r Theoretische Physik und Astrophysik, \\ Universit{\"a}t W{\"u}rzburg,
       97074 W{\"u}rzburg, Germany

\end{center}}

\vspace*{1.5cm}

\begin{center}
{\Large \today}
\end{center}

{\Large \bf
\begin{center} Abstract \end{center}  }

We discuss the interplay between spectral shape and detector response beyond a simple $E^{-2}$ neutrino flux at neutrino telescopes, at the example of time-integrated point source searches using IceCube-40 data. We use a self-consistent model  for the neutrino production, in which protons interact with synchrotron photons from co-accelerated electrons, and we fully take into account the relevant pion and kaon production modes, the flavor composition at the source, flavor mixing, and magnetic field effects on the secondaries (pions, muon, and kaons). Since some of the model parameters can be related to the Hillas parameters $R$ (size of the acceleration region) and $B$ (magnetic field), we relate the detector response to the Hillas plane. In order to compare the response to different spectral shapes, we use the energy flux density as a measure for the pion production efficiency times luminosity of the source. We demonstrate that IceCube has a very good reach in this quantity for AGN nuclei and jets for all source declinations, while the spectra of sources with strong magnetic fields are found outside the optimal reach. We also demonstrate where neutrinos from kaon decays and muon tracks from $\tau$ decays can be relevant for the detector response. Finally, we point out the complementarity between IceCube and other experiments sensitive to high-energy neutrinos, at the example of 2004--2008 Earth-skimming neutrino data from Auger. We illustrate that Auger, in principle, is better sensitive to the parameter region in the Hillas plane from which the highest-energetic cosmic rays may be expected in this model.

\vspace*{.5cm}

\end{titlepage}

\newpage

\renewcommand{\thefootnote}{\arabic{footnote}}
\setcounter{footnote}{0}

\section{Introduction}

Neutrino telescopes, such as IceCube~\cite{Ahrens:2002dv} or ANTARES~\cite{Aslanides:1999vq}, are designed to detect neutrinos from astrophysical sources. If protons are accelerated in these astrophysical objects, as we expect from the observation of the highest energetic cosmic rays, the collision with target photons or protons will lead to charged pion production, and therefore to an extraterrestrial neutrino flux.
 Therefore, neutrino telescopes are an indirect method to search for the origin of the cosmic rays. 
 There are numerous source candidates, the most prominent extragalactic ones being Active Galactic Nuclei (AGNs)~\cite{Stecker:1991vm,Mannheim:1993,Mucke:2000rn,Aharonian:2002} and Gamma-Ray Bursts (GRBs)~\cite{Waxman:1997ti}, see \Ref~\cite{Becker:2007sv} for a review and \Ref~\cite{Rachen:1998fd} for the general theory of the astrophysical neutrino sources. In generic estimates or bounds, such as \Ref~\cite{Waxman:1998yy,Mannheim:1998wp}, the cosmic ray flux is related to the potentially expected neutrino flux. Very interestingly, IceCube-40, referring to the 40 string configuration of IceCube, is currently starting to touch these generic estimates for particular source candidates, see \Ref~\cite{Abbasi:2011qc} for GRBs. In addition, time-integrated~\cite{Abbasi:2010rd} and time-dependent~\cite{Abbasi:2011ara} point source searches have been performed, so far, without success. On the other hand, there may be sources for which the optical counterpart is absorbed, so-called ``hidden sources'', see, \eg, \Ref~\cite{Razzaque:2004yv,Ando:2005xi,Razzaque:2005bh,Razzaque:2009kq}. 
This immediately raises questions how to interpret the data (see also \Ref~\cite{Arguelles:2010yj} for AGN models), in particular: what does it mean that IceCube has not seen anything? What parts of the parameter space is IceCube actually most sensitive to?
In this study, we address these questions in terms of the interplay between spectral shape expected from the sources and the detector response. In order to quantify the detector response, we use the time-integrated point source analysis in \Ref~\cite{Abbasi:2010rd} for different source declinations.  In addition, we compare the parameter space coverage to other experiments, at the example of 2004--2008 Earth-skimming neutrino data from Auger~\cite{Abraham:2009uy}.

\begin{figure}[t]
\begin{center}
\includegraphics[width=0.45\textwidth]{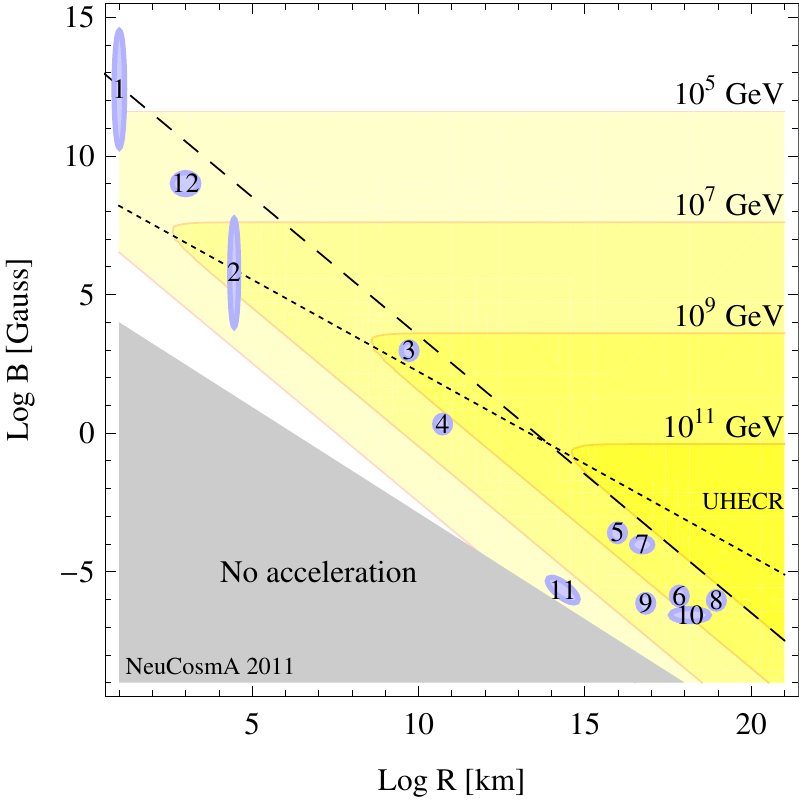}
\end{center}
\mycaption{\label{fig:maxp} Maximal proton energy as a function of $R$ and $B$ in our model (contours) for an acceleration efficiency $\eta=0.1$ on the Hillas plot; data taken from Fig.~2 in \Ref~\cite{Hummer:2010ai}. 
The dashed line indicates the Hillas condition in \equ{hillas} for $10^{20} \, \mathrm{eV}$ protons. The dotted line  separates the regions where synchrotron (above line) and adiabatic (below line) energy losses dominate the maximal proton energy in the model, \ie, the maximal proton energy follows the Hillas condition below the dotted line. The region ``UHECR'' indicates where $10^{20} \, \mathrm{eV}$ cosmic ray protons are expected to be produced in the model. The ``no acceleration'' region refers to either inefficient acceleration, or inefficient pion production. The different ``test points'' (numbered disks) correspond to: (1) neutron stars, (2) white dwarfs, (3) active galaxies: nuclei, (4) active galaxies: jets, (5) active galaxies: hot-spots, (6) active galaxies: lobes, (7) colliding galaxies, (8) clusters, (9) galactic disk, (10) galactic halo, (11) supernova remnants, (12) additional test point. Test points taken from \Ref~\cite{Bertou:2000ip}.
 }
\end{figure}

A convenient description of the parameter space of interest is the Hillas plot~\cite{Hillas:1985is}. In order to confine a particle in a magnetic field at the source, the Larmor radius has to be smaller than the extension of the acceleration region $R$. This can be translated into the Hillas condition for the maximal particle energy
\begin{equation}
E_{\mathrm{max}} \, [\mathrm{GeV}] \simeq 0.03 \cdot \eta \cdot Z \cdot  R \, [\mathrm{km}] \cdot B \, [\mathrm{G}] \, .
\label{equ:hillas}
\end{equation}
Here $Z$ is the charge (number of unit charges) of the accelerated particle, $B$ is the magnetic field in Gauss, and $\eta$ can be interpreted as an efficiency factor or linked to the characteristic velocity of the scattering centers. Potential cosmic ray sources are then often shown in a plot as a function of $R$ and $B$, as it is illustrated in \figu{maxp} by the numbered disks (see end of figure caption for possible source correspondences). Assuming that a source produces the highest energetic cosmic rays with $E \simeq 10^{20} \, \mathrm{eV}$, one can interpret \equ{hillas} as a necessary condition excluding the region below the dashed line in \figu{maxp} (for protons with $\eta=0.1$). However, this method does not take into account energy loss mechanisms, which may lead to a qualitatively different picture -- as we shall see later. In the following, we will study the complete parameter space covered by \figu{maxp} without any prejudice. Since the location of the sources in \figu{maxp} cannot be taken for granted, we will refer to the individual sources in \figu{maxp} as ``test points'' (TP) in most cases, and leave the actual interpretation to the reader.

Our main focus is the particle physics perspective, \ie, we start off with the minimal ingredients to neutrino production and the main impact factors to the spectral shape. First of all, note that most neutrino telescope analyses (and even some models) use an $E^{-2}$ neutrino spectrum as an initial assumption, see, \eg, \Ref~\cite{Abbasi:2010rd}, which is often believed to be consistent with a  proton injection spectrum ($\propto E^{-2}$) coming from Fermi shock acceleration. However, in $p\gamma$ interactions, the neutrino spectrum follows the pion spectrum, which depends on proton {\em and} photon spectral shape. Therefore, the $E^{-2}$ assumption for the neutrinos only holds for an $E^{-1}$ target photon density (for the $E^{-2}$ proton injection spectrum), as it is often assumed for the prompt emission from GRBs up to the photon break. In self-consistent models, where the target photons originate, for example, from synchrotron emission of co-accelerated electrons, such a hard spectrum is very difficult to obtain, which results in a different neutrino spectrum; see, \eg, \Ref~\cite{Mucke:2000rn} for an example. In this work, we study the detector response to different shapes of the neutrino spectra. For that purpose, we require a toy model which can predict the neutrino spectral shape for wide regions of the parameter space without any bias from a multi-messenger observation, since, after all, neutrino sources may not be necessarily seen as photon sources. The model needs to take into account the dominant particle physics processes especially affecting the spectral shape, which are multi-pion processes in the pion production (see, \eg, \Refs~\cite{Murase:2005hy,Lipari:2007su,Hummer:2010vx}), neutron and kaon production (see, \eg, \Refs~\cite{Kachelriess:2006fi,Asano:2006zzb,Baerwald:2010fk,Hummer:2010ai}), magnetic field effects on the secondary muons, pions, and kaons (see, \eg, \Refs~\cite{Kashti:2005qa,Kachelriess:2007tr,Lipari:2007su,Reynoso:2008gs,Baerwald:2010fk,Hummer:2010ai}), and the appropriate maximal energies. Note that for the neutrino spectral shape, especially the magnetic field is an important control parameter, see, \eg, \Refs~\cite{Hummer:2010ai,Baerwald:2011ee} for quantitative discussions. First of all, the maximal proton energy, which is recovered in the neutrino spectrum, can be limited by proton synchrotron emission. Second, for high enough magnetic fields, both the maximal energy of the neutrino spectrum and the spectral shape will be determined by the synchrotron cooling of the secondaries. Therefore, in order to predict the neutrino spectral shape, it is, especially for large $B$, more important to accurately model the particle physics of the secondaries instead of the cooling and escape processes of the primaries (protons, electrons, photons). Especially if the target photons come from synchrotron emission, there is little sensitivity to the spectral shape of the parents (only the square root of the parent spectral index enters the photon spectrum).

We use the model in \Ref~\cite{Hummer:2010ai} for the prediction of the spectral shapes, where one of the starting points was the minimal set of assumptions required to describe the neutrino flavor ratios and spectral shapes as accurate as possible, given the above boundary conditions. In this model, charged pions are produced from photohadronic ($p\gamma$) interactions  between protons and the synchrotron photons from co-accelerated electrons (positrons). The photohadronic interactions are computed using the method described in \Ref~\cite{Hummer:2010vx}, based
on the physics of SOPHIA~\cite{Mucke:1999yb}, including higher resonances, $t$-channel charged pion production, and multi-pion production. The helicity-dependent muon decays are taken into account as described in \Ref~\cite{Lipari:2007su}. The toy model relies on relatively few astrophysical parameters, the most
                important ones  being  the size of the acceleration region ($R$), the magnetic field strength at the source ($B$),
  and  the injection index ($\alpha$) which is assumed to be universal for protons and electrons/positrons.
 Naturally, the parameters $R$ and $B$ can be directly related to the Hillas plot. 
The leading kaon production mode and the energy losses of all secondaries (muons, pions, kaons) are taken into account. Our key argument goes as follows: for each set of $R$, $B$, and $\alpha$, we can predict the spectral flux shape using this model. Using the exposures from IceCube or Auger, we can then compute  the sensitivity limit (normalization of the flux) for exactly this parameter set. In order to compare sensitivities to different spectra, the (integrated) neutrino energy flux density at the detector [$\mathrm{erg} \, \mathrm{cm^{-2}} \, \mathrm{s^{-1}}$] is used. We illustrate at the end of \Sec~\ref{sec:method}, that this quantity is typically a direct measure for the luminosity $\times$ pion production efficiency of the source. Thus, ``good sensitivity'' in this work means that the source will be found for relatively small values of this product. 
Note that we do not interpret the data in terms of the deeper astrophysics involved, and we do not predict the absolute levels of the neutrino fluxes. In addition, note that our conclusions will naturally be somewhat model-dependent. However, as we will illustrate, some of the qualitative results should be rather robust. 

This study is organized as follows: In \Sec~\ref{sec:model}, we summarize the key ingredients of the model, where a critical discussion of the limitations of the model can be found at the end of this section.  Then in \Sec~\ref{sec:method}, we describe our method to limit individual fluxes using the neutrino effective areas from IceCube-40. Furthermore, in \Sec~\ref{sec:constraints} we show the result for the whole Hillas plane and different source declinations, we illustrate the impact of the injection index $\alpha$, neutrinos from kaon decays, and $\nu_\tau$-induced muon tracks. In \Sec~\ref{sec:compl}, we emphasize the complementarity of different experiments and data sets, using Auger as an additional example. Finally, we summarize in \Sec~\ref{sec:summary}.

\section{Review of the source model}
\label{sec:model}

The model in \Ref~\cite{Hummer:2010ai} describes neutrino production via photohadronic ($p\gamma$) processes for transparent sources (optically thin to neutrons) and includes magnetic field effects on the secondary particles (pions, muons, kaons). It can be used to generate
neutrino fluxes as a function of few astrophysical parameters. Below we outline the key ingredients of the
model relevant for this study; for details  see \Refs~\cite{Hummer:2010vx,Hummer:2010ai}, or \Ref~\cite{Mehta:2011qb} for a shorter summary. All of the following
quantities refer to the frame where the target photon field is isotropic, such as the shock rest frame (SRF).

The protons and electrons/positrons are injected with spectra $\propto E^{-\alpha}$, where $\alpha$ is one of the main model parameters. The maximal energies of
these spectra are determined by balancing the energy loss and acceleration timescale given by
  \begin{equation}
  t^{-1}_\mathrm{acc}=\eta\frac{c^2 e B}{E} \,,
 \end{equation}
with $\eta$ an acceleration efficiency depending on the acceleration mechanism, where we typically choose $\eta=0.1$. If synchrotron losses dominate, the maximal energy is therefore given by
\begin{equation}
E_{\mathrm{max}} = \sqrt{ \frac{9 \pi \epsilon_0 \eta}{B}} \frac{m^2 c^{7/2}}{e^{3/2}} \, .
\end{equation}
It scales $\propto m^2$, which means that the protons are accelerated to much higher energies, and $\propto 1/\sqrt{B}$, which means that strong magnetic fields limit the maximal energies. If adiabatic energy losses dominate, $t^{-1}_{\mathrm{ad}} \simeq c/R$, the maximal energy is (for protons) given by \equ{hillas}, \ie, the Hillas condition. Since the neutrino energy follows the proton energy, the maximal energy of the protons determines (for not too strong magnetic field effects) the peak of the neutrino spectrum in $E^2 dN/dE$. 
We show the maximal proton energy in \figu{maxp} as a function of $R$ and $B$ for this model, where the dotted line separates the regions where synchrotron (above line) and adiabatic (below line) energy losses dominate the maximal proton energy (here $\eta=0.1$).\footnote{Note that these assumptions are consistent with \Ref~\cite{Medvedev:2003sx} (efficient acceleration case) and \Ref~\cite{Protheroe:2004rt}, who have emphasized these effects earlier. Especially, the dotted line approximately corresponds to the middle solid curve in Fig.~6 of \Ref~\cite{Protheroe:2004rt}, which limits the region where the Hillas condition for the maximal proton energy applies. That, however, does not mean that the region on the r.h.s. of the dotted line is excluded, it only means that the maximal energy is limited otherwise (by synchrotron losses, and, in the very lower right corner of our plot, also by interactions with the CMB). Compared to \Ref~\cite{Protheroe:2004rt}, we show in \figu{maxp} the maximal energy directly. }
 One can easily see that in the upper part (above the dotted line) the Hillas condition, \equ{hillas}, does not apply, since the synchrotron losses dominate, whereas below the dotted line, the maximal proton energy follows \equ{hillas}. One can also get an educated guess for the best IceCube spectral shape sensitivity already, since its differential limits are minimal at around $10^4$ to $10^6 \, \mathrm{GeV}$ (see \Sec~\ref{sec:method}) -- which should be a very robust prediction. In the absence of magnetic field effects, one can estimate that the optimal detector response is roughly obtained when these differential limit minima coincide with a certain fraction of the maximal proton energy in \figu{maxp} (the neutrinos take about 0.05 to 0.1 of the proton energy in photohadronic interactions). Note that in this model, the potential sources of the highest-energetic cosmic ray protons are to be found in the lower right corner in the region labeled ``UHECR'' within the $10^{11} \, \mathrm{GeV}$ contour (unless there is a  strong Lorentz boost of the source). In this case, the neutrino energies extend up to about $10^{10} \, \mathrm{GeV}$. However, recent results on the cosmic ray composition, such as from Auger~\cite{Abraham:2010yv}, indicate that, at least above about $3 \, 10^{9}$~GeV, heavier nuclei may dominate the cosmic ray composition. Since these heavier nuclei may reach the same energies for lower magnetic fields according to \equ{hillas}, they would occupy a slightly different region in the Hillas plane. Therefore, one should keep in mind that the mapping into the Hillas plane only exactly applies to the proton contribution at ultra-high energies.

For each particle species, the injection and energy losses/escape are balanced by the steady state equation
\begin{equation}
\label{equ:steadstate}
Q(E)=\frac{\partial}{\partial E}\left(b(E) \, N(E)\right)+\frac{N(E)}{t_\mathrm{esc}} \, ,
\end{equation}
with $t_\mathrm{esc}(E)$ the characteristic escape time, $b(E)=-E\,t_\mathrm{loss}^{-1}$ with
$t_\mathrm{loss}^{-1}(E)=-1/E \, dE/dt$ the rate characterizing energy losses, $Q(E)$ the particle injection
rate $[\mathrm{\left( GeV\,s\,cm^3\right)^{-1}}]$ and $N(E)$ the steady particle spectrum $[\mathrm{\left(
GeV\,cm^3\right)^{-1}}]$. For all charged particles, synchrotron energy losses  and adiabatic cooling  are
taken into account.  In addition, unstable secondaries, \ie, pions, muons, and kaons, may escape via decay. As a
consequence, for pions, muons, and kaons, neglecting the adiabatic cooling, the (steady state) spectrum is
loss-steepened above the energy
\begin{equation}
E_c = \sqrt{ \frac{9 \pi \epsilon_0}{\tau_0}} \frac{m^{5/2} c^{7/2}}{e^2 B} \, ,
\label{equ:ec}
\end{equation}
where synchrotron cooling and decay rates are equal. One can read off this formula that the different
secondaries, which have different masses $m$ and rest frame lifetimes $\tau_0$,  will exhibit different break
energies $E_c \propto \sqrt{m^5/\tau_0}$ which solely depend on particle physics properties and the value of
$B$. These different break energies will lead to a spectral split of the neutrino spectra, which is an imprint of the magnetic field. 

While being accelerated, the electrons loose energy into synchrotron photons, which serve as the target
photon field. Charged meson production then occurs via
\begin{align}
 p + \gamma & \rightarrow \pi + p' \label{equ:photo1} \, ,\\ 
 p + \gamma & \rightarrow K^+ + \Lambda/\Sigma \, , \label{equ:photo2}
\end{align}
with these synchrotron photons, where the leading kaon production mode is included and $p'$ is a  proton or
neutron.  In addition, two- and multi-pion production processes are included (not listed here), see \Ref~\cite{Hummer:2010vx}
for details. The injection of the charged mesons is computed from the steady state proton $N_p(E_p)$ and photon
$N_\gamma(\varepsilon)$ spectra with~\Ref~\cite{Hummer:2010vx}
\begin{equation}
Q_b(E_b) = \int\limits_{E_b}^{\infty} \frac{dE_p}{E_p} \, N_p(E_p) \, \int\limits_{0}^{\infty} d\varepsilon \, N_\gamma(\varepsilon) \,  R_b( x,y ) \,,
\label{equ:prodmaster}
\end{equation}
with $x=E_b/E_p$ the fraction of energy going into the secondary, $y \equiv (E_p\varepsilon)/m_p$  (directly
related to the center of mass energy) and a ``response function'' $R_b( x,y )$ (see \Ref~\cite{Hummer:2010vx}). The weak decays of the secondary mesons, such as 
\begin{eqnarray}
\pi^+ & \rightarrow & \mu^+ + \nu_\mu
 \, , \label{equ:piplusdec} \\
& & \mu^+ \rightarrow e^+ + \nu_e  + \bar\nu_\mu   \label{equ:muplusdec} \, ,
\end{eqnarray}
are described in \Ref~\cite{Lipari:2007su}, including the helicity dependence of the muon decays. These will finally lead to neutrino fluxes from pion, muon, kaon,
and neutron decays. 

In order to compute the $\nu_\beta$ ($\beta=e$, $\mu$, $\tau$) neutrino flux at the detector, \ie, including flavor mixing, we sum over all these initial neutrino fluxes of flavor $\nu_\alpha$ weighted by the usual flavor mixing
\begin{equation}
 P_{\alpha \beta} = \sum\limits_{i=1}^3 |U_{\alpha i}|^2 | U_{\beta i}|^2 \, ,
\end{equation}
where $U_{\alpha i}$ are the entries of the Pontecorvo-Maki-Nakagawa-Sakata mixing matrix. We use $\sin^2 \theta_{12}=0.318$, $\sin^2 \theta_{23}=0.5$, and $\sin^2 \theta_{13}=0$ for the sake of simplicity (see, \eg, \Ref~\cite{Schwetz:2008er}), leading to flavor equipartition between $\nu_\mu$ and $\nu_\tau$ at the detector. We sum over neutrinos and antineutrinos, \ie,  if we refer to ``$\nu_\mu$'', we mean the sum of the $\nu_\mu$ and $\bar\nu_\mu$ fluxes.

Concerning the limitations of the model, it certainly does not apply exactly to all types of sources.  
For example, in supernova remnants, $pp$ (proton-proton) or $pA$ (proton-nucleus) interactions may dominate the neutrino production, which would require additional parameters to describe the target protons or nucleons. In addition, at ultra-high energies, heavier nuclei may be accelerated, see discussion above. 
The spirit of this model is different:  It is developed as  the simplest (minimal) possibility including non-trivial magnetic field and flavor effects. The relevant point for this study is that the instrument response depends on the neutrino energies and spectral features. The spectral features, as
a peculiarity of neutrinos, are in cases of strong magnetic fields dominated by the cooling and decay of the secondaries (pions, muons, kaons). The energy of the secondaries follows the proton energy, to a first approximation, $E_b \sim x \, E_p$ with $x = \mathcal{O}(0.1)$ to $\mathcal{O}(1)$. These two observations apply to any of these mentioned interactions, no matter if $p\gamma$, $pp$, or $pA$, if the energy of the accelerated particle is much larger than the one of the interaction partner (Feynman scaling). Therefore, we expect qualitatively similar results for these cases.

Another variable is the  target photon density, which is assumed to come from synchrotron emission of co-accelerated electrons here. In more realistic models, typically a combination of different radiation processes is at work. For example, there may be thermal photons radiated off an accretion disk, proton synchrotron photons, inverse Compton up-scattered photons, photons from pair annihilation, and photons from $\pi^0$ decays cascaded down to lower energies. It is therefore a frequently used approach for a particular type of source to compute the neutrino spectrum from a target photon spectrum which corresponds to the observation without describing the origin of these photons, see, \eg, \Ref~\cite{Abbasi:2011qc} for gamma-ray bursts, where a band function parameterization  of the observed gamma-ray fluxes is used.
Qualitatively,  the target photons control the interaction rate of the primaries in $p\gamma$ interactions [see, \eg, \cite{Hummer:2010vx}], which enters the normalization and is not used here. In addition, the spectral shape depends on the target photon spectrum in the energy range relevant for the interactions.  
In many examples with strong magnetic fields a spectral break in the photon spectrum is less  important than the cooling and decay of the secondaries, which depend on particle physics only. For example, in our model, different hypotheses for the cooling of the electrons, leading to the target photons, have been tested, and in most cases the relevant part of the spectral shape hardly changes. Thus, while it is unlikely that the model applies exactly to a particular source, it may be used as a good starting hypothesis. In addition, note that from \equ{prodmaster}, only the product of the proton and photon  (and therefore electron) density normalizations enters the final result. The final normalization of the neutrino spectrum will depend on the source luminosity, the interaction volume, a possible Lorentz boost of the acceleration region $\Gamma$, and the redshift $z$ of the source. Since we only need the spectral shape, we do not compute the normalization explicitely. Note, however, that also the final neutrino energies depend on $\Gamma$ and $z$. For the sake of simplicity, we neglect these effects, but one should keep in mind that the actually observed energy spectrum could be significantly shifted in energy  as $E \propto \Gamma/(1+z)$. 

The main parameters of the model are $R$, affecting the shape and maximal energy (via $t^{-1}_{\mathrm{ad}}$) of the primaries, $B$, affecting the maximal energy (via $t^{-1}_{\mathrm{synchr}}$) of the primaries and the break (via $E_c$) of the secondaries, and $\alpha$, affecting the spectral slope of the primaries. The parameters $R$ and $B$ can be directly related to the Hillas parameters, see \figu{maxp}.

\section{Method and impact of the spectral shape}
\label{sec:method}

Here we describe our method, \ie, how we constrain individual fluxes from IceCube data and quantify the response of the instrument. The simplest possible approach is total event rate-based, which means that no information on the reconstructed neutrino or muon energy is used explicitely. It can be described with the exposure $\mathrm{Exp}(E,\delta) \equiv A_\nu^{\mathrm{eff}}(E,\delta) \, t_{\mathrm{exp}}$, where $A_\nu^{\mathrm{eff}}$ is the neutrino effective area and $t_{\mathrm{exp}}$ is the observation time. Here  $A_\nu^{\mathrm{eff}}(E,\delta)$ is a function of the flavor or interaction type (which we do not show explicitely), the incident neutrino energy $E$, and the declination of the source $\delta$. The neutrino effective area already includes Earth attenuation effects and event selection cuts to reduce the backgrounds, which depend on the type of source considered, the declination, and the assumptions for the input neutrino flux, such as the spectral shape. Normally, the cuts are optimized for an $E^{-2}$ flux, which means that for specific fluxes with different shapes, the following analysis may slightly improve by an optimization of the detector response. On the other hand, one has to understand that the experiments cannot optimize their event selection for any possible input spectrum.
The total event rate of a neutrino telescope can be obtained by folding the input neutrino flux with the exposure as
\begin{equation}
 N = \int dE  \, \mathrm{Exp}(E,\delta) \, \frac{dN(E)}{dE}= \int dE A_\nu^{\mathrm{eff}}(E, \delta) \, t_{\mathrm{exp}} \, \frac{dN(E)}{dE}  \, .
\label{equ:N}
\end{equation} 
Here $dN(E)/dE$ is, for point sources, given in units of $\mathrm{GeV^{-1} \,  cm^{-2} \, s^{-1}}$. If backgrounds are negligible, the 90\% (Feldman-Cousins) sensitivity limit $K_{\mathrm{90}}$ for an arbitrarily normalized input flux used in \equ{N} can be estimated as $K_{\mathrm{90}} \sim 2.44/N$~\cite{Feldman:1997qc}. This imples that a predicted flux at the level of the sensitivity limit, irrespective of the spectral shape, would lead to the same number (2.44) of events. The 90\% confidence level differential limit $E^2 dN/dE$ can be defined as $2.3 \, E/\mathrm{Exp}(E,\delta)$, see, \eg, \Ref~\cite{Abraham:2009uy}.\footnote{This is the factor weighting $E^2 dN/dE$ in the integrand of \equ{N} if integration over $\mathrm{Log}(E)$ is used. If the differential limit and flux are smooth enough on a logarithmic energy  scale, the flux limit typically is below the differential limit since enough contribution to the integral is obtained. However, if spikes or sudden flux or differential limit changes are present, the flux limit may also exceed the differential limit locally. An example is the Glashow resonance process, which is sometimes taken out of the analysis for this reason. Note that in this study, $\mathrm{Log}(E)=\mathrm{Log}_{10}(E)$ everywhere.} We will comment on the impact of the backgrounds below.

\begin{figure}[tp]
\begin{center}
\includegraphics[width=0.85\textwidth]{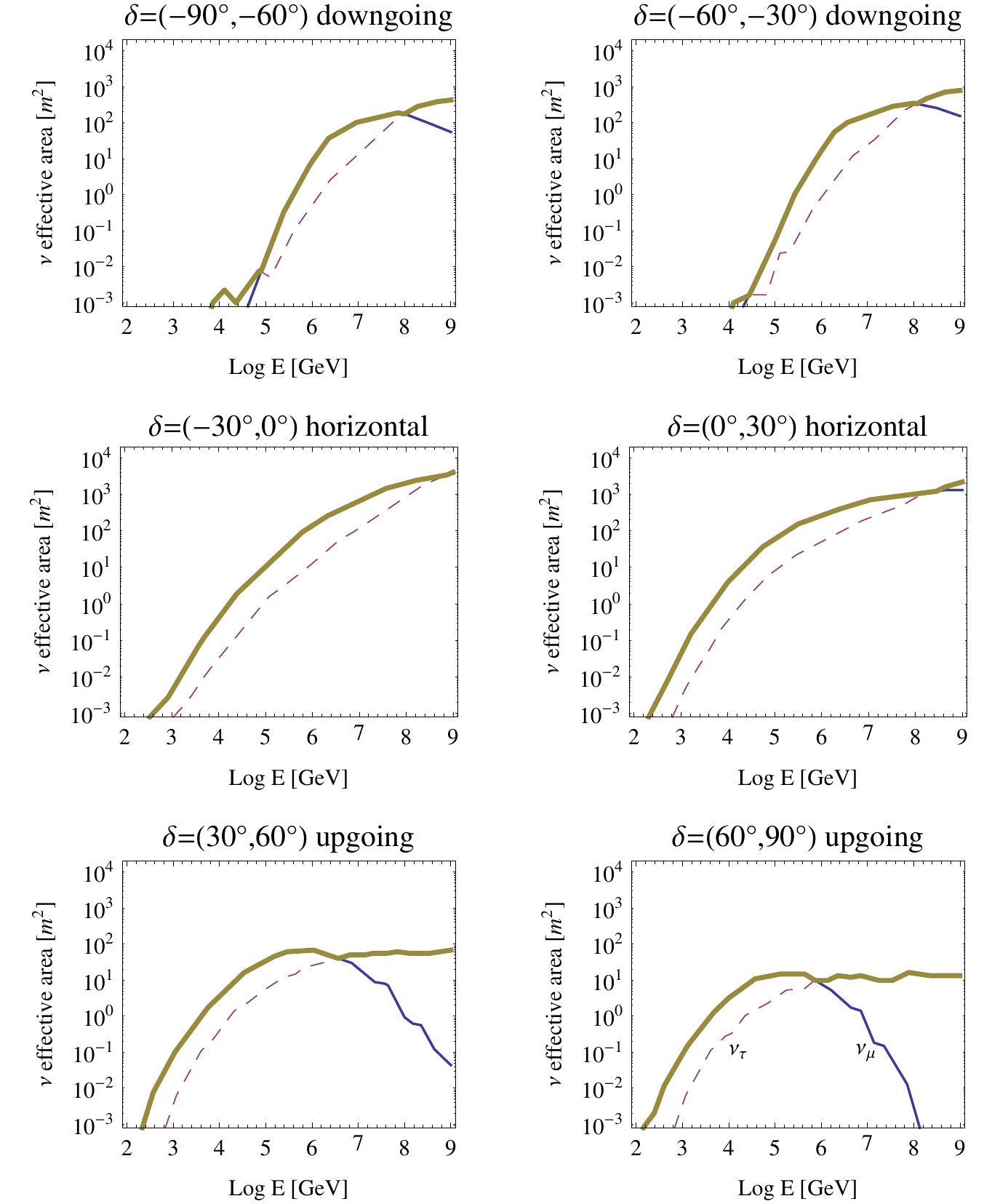}
\end{center}
\mycaption{\label{fig:aeff} Neutrino effective area as a function of the neutrino energy for different declinations in IceCube-40. The thin solid curves correspond to muon tracks from $\nu_\mu$, the dashed curves to muon tracks from $\nu_\tau$ after $\tau$ decay. The thick solid curves show the neutrino effective area for either $\nu_\mu$ or $\nu_\tau$ if flavor equipartition between $\nu_\mu$ and $\nu_\tau$ is assumed. Data taken from Fig.~8 in \Ref~\cite{Abbasi:2010rd} and re-arranged. }
\end{figure}

In the following, we use the neutrino effective areas for time-integrated point source searches in IceCube-40~\cite{Abbasi:2010rd}, based on 2008--2009 data with  $t_{\mathrm{exp}}=375.5$ days. We show these neutrino effective areas for muon tracks and for six declination bands, as in the original reference, in \figu{aeff}.   For the sake of readability, we use ``downgoing'', ``upgoing'', or ``(quasi-)horizontal'' as additional description of the declination bands. Note that for IceCube located at the South Pole, a particular source will always appear under a specific declination, which means that the position in the sky will have some impact on the sensitivity, and the declination bands correspond to different source classes or data sets.
Since upgoing muon neutrinos become absorbed in Earth matter above about 30~PeV and strong cuts have to be applied for the downgoing events to reduce the cosmic muon background,  quasi-horizontal
events imply the optimal performance.

The effective areas shown in \figu{aeff} are for muon tracks only. However, note that there are two contributions: not only $\nu_\mu$ produce muon tracks, but also $\nu_\tau$, since the tau lepton decays into a muon with a branching ratio of about 17\%.
We show the individual contributions in \figu{aeff}, where thin solid curves represent the $\nu_\mu$ effective areas, the dashed curves the $\nu_\tau$ effective areas, and the thick solid curves the total effective areas for either $\nu_\mu$ or $\nu_\tau$ if flavor equipartition $\nu_\mu$:$\nu_\tau$=1:1 is assumed. Note that the total effective areas for the individual  $\nu_\mu$ or $\nu_\tau$ fluxes are given by the maximum of the individual effective areas, not the sum. Assuming flavor equipartition, one can read off from \figu{aeff} that quasi-horizontal muon tracks are dominated by $\nu_\mu$ events. The downgoing muon tracks also come mostly from $\nu_\mu$, but there is a small region at high energies where $\nu_\tau$ dominate. In this case, the muon tracks are recovered at lower energies. The upgoing muon tracks above 30~PeV, where the Earth becomes opaque to $\nu_\mu$, are dominated by $\nu_\tau$ events. Therefore, it is {\em a priori} not clear if and when the muon tracks from $\nu_\tau$ can be neglected. In general, our analysis will be based on $\nu_\mu$ interactions, but we will come back to this point in \Sec~\ref{sec:nutau}. Note that beyond the limits shown in \figu{aeff}, we use linear extrapolation, which turns out to be a good approximation because it only effects the regions of poor sensitivity.

The backgrounds for the point source analysis in \Ref~\cite{Abbasi:2010rd} mostly come from atmospheric neutrinos for upgoing events, and from cosmic muons for downgoing events. For our analysis, we assume that backgrounds can be suppressed to a sufficient level, which, for point sources, can be achieved by the angular resolution of about one degree for upgoing events. For example, if the photon counterpart is used to pre-select potential sources (``{\em a priori} source candidate search''), the number of background events is relatively low, see Table~3 in \Ref~\cite{Abbasi:2010rd}. For downgoing events, the suppression of the cosmic-ray muon background is more difficult, which means that here the assumption of low backgrounds has to be interpreted with care and somewhat higher backgrounds are expected.\footnote{In fact, the cuts for the downgoing events seem to be chosen such that the backgrounds are roughly a factor of two higher for the downgoing events than the upgoing events, see examples in Table~3 in \Ref~\cite{Abbasi:2010rd}. However, the absolute number of expected background events is nevertheless only a few, which means that the impact on the final sensitivity limit using the Feldman-Cousins approach~\cite{Feldman:1997qc} is relatively small on a logarithmic scale.}
In either case, we can reproduce the sensitivity for an $E^{-2}$ flux in Fig.~19 of \Ref~\cite{Abbasi:2010rd} very well. We have also checked the flux of atmospheric neutrinos to be expected within one degree angular resolution, see, \eg, \Ref~\cite{Abbasi:2009nfa}. It turns out that especially in the energy range between 100~GeV and 10~TeV, where Earth attenuation effects are relatively small, the  IceCube-40 point source sensitivities for upgoing events are already touching the atmospheric neutrino background in the most conservative (systematics) case for the atmospheric neutrino flux. This means that in this energy range, we expect that the IceCube sensitivity will be background-limited very soon, whereas for the present analysis, neglecting the atmospheric neutrino background is a good approximation.

\begin{figure}[tp]
\begin{center}
\includegraphics[width=0.9\textwidth]{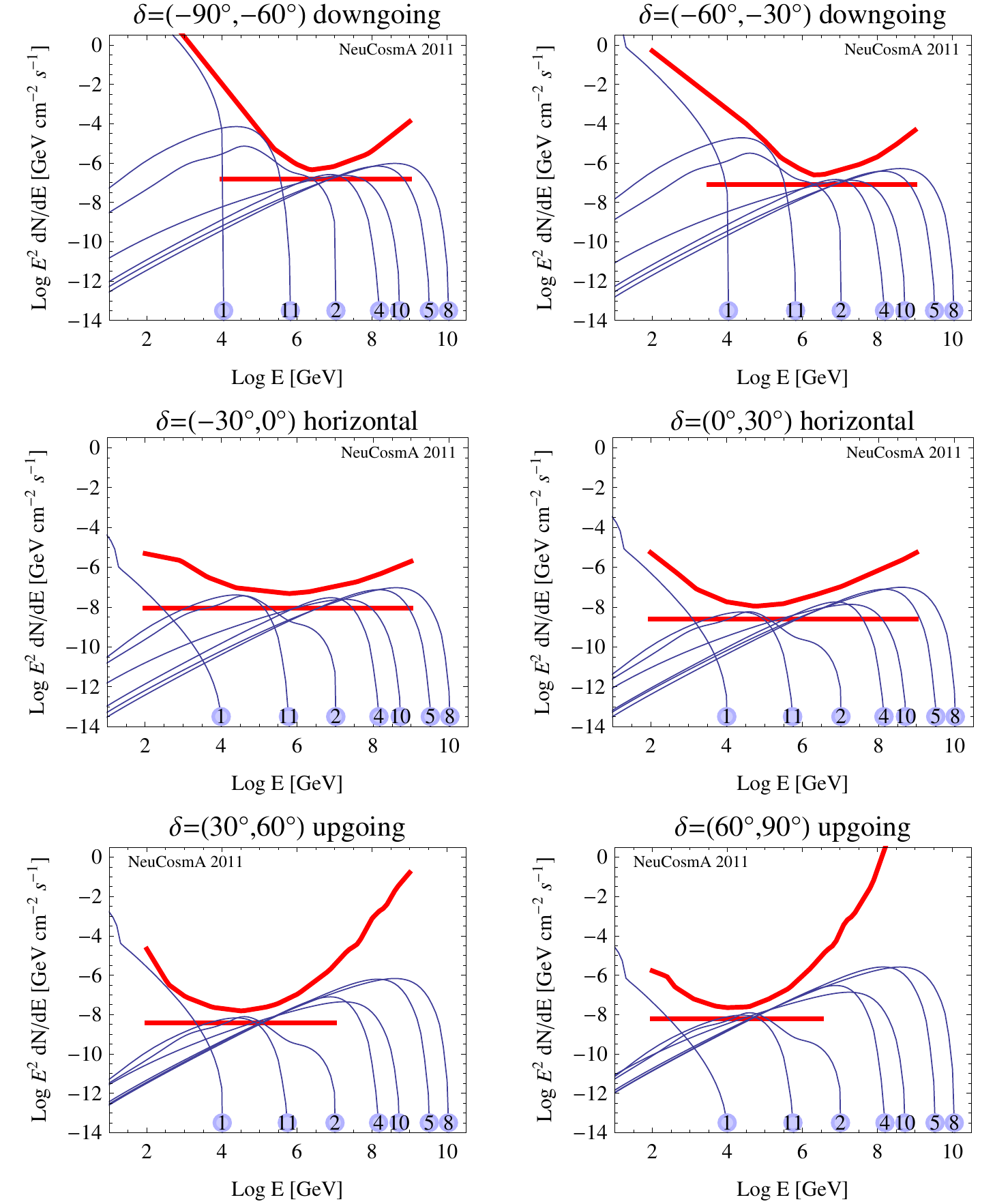}
\end{center}
\mycaption{\label{fig:spectra} Limits for selected $\nu_\mu$ spectra (including flavor mixing) for different declination bands in IceCube-40 (90\% CL). The numbers of the individual curves correspond to the test points in \figu{maxp}. The thick lines show the limits for an $E^{-2}$ flux (in the dominant energy range), and the thick curves the differential limits. Here the injection index $\alpha=2$ is chosen. }
\end{figure}

We apply the above mechanism to obtain a limit for arbitrary fluxes to our model. We show in \figu{spectra} the limits for selected $\nu_\mu$ spectra (including flavor mixing) for different declination bands in IceCube-40. We also show as thick lines the limits for an $E^{-2}$ flux (in the dominant energy range\footnote{The choice of this energy range is somewhat arbitrary, as long as the main contribution to the integral, \ie, the minimum of the different limit is well contained in the integration range. Often the range where 90\% of the events are expected is shown.}), and as thick curves the differential limits. The $\nu_\mu$ spectra correspond to the test points as marked in the figure, which correspond to the parameters $R$ and $B$ in \figu{maxp} and $\alpha=2$. From the differential limits, we can read off that the optimum sensitivity for downgoing events is found at higher energies $\sim$PeV, whereas upgoing events are best constrained at about 10~TeV~\cite{Abbasi:2010rd}. The quasi-horizontal events exhibit a relatively broad differential limit. As far as the absolute sensitivity is concerned (see, \eg, thick solid lines for integrated limits), horizontal and upgoing events have similar sensitivities, but the downgoing events face lower statistics because of the necessary cuts to reduce the cosmic-ray muon background.  As far as the limits for the individual spectra are concerned, we make a few observations. First of all, the best sensitivity, \ie, lowest normalization, for a particular spectrum is typically obtained if the spectral peak in $E^2 dN/dE$ coincides with the minimum of the differential limit. The optimum sensitivity is then, to a first approximation, given by the maximum proton energy, which roughly determines the position of the peak and can be read off from \figu{maxp}. For example, TP~11 is best constrained by upgoing events. However, especially if strong magnetic field effects are present and therefore the spectral shape becomes more complicated, such as for TP~2 for downgoing events, this rule does not necessarily apply anymore. In this case, there can be quite some impact of neutrinos from kaon decays, as we will discuss in  \Sec~\ref{sec:kaon}. Note that, for a particular source, the declination is, of course, pre-determined. The way to interpret \figu{spectra} therefore goes as follows: if a particular spectral shape for a source of declination $\delta$ is described by a TP of this model, the sensitivity can be read off from the corresponding panel of \figu{spectra}.

From \figu{spectra}, one can easily see that it is not trivial how to compare two different spectra with different spectral shapes. Consider, for example, the fluxes~2 and~11 in the upper left panel, both leading to the same event rate by definition.  Which of the two neutrino sources leading to these fluxes can be better constrained by the downgoing events? In order to quantify this aspect, it is useful to assign a single number to each spectrum which measures how much energy in neutrinos can be tested for a specific spectrum and event type. We choose the energy flux density
\begin{equation}
\phi = \int E \frac{dN(E)}{dE} d E \,  \label{equ:eflux}
\end{equation}
as this quantity, which we show in units of $\mathrm{erg \, cm^{-2} \, s^{-1}}$ for point sources in order to distinguish it from $E^2 dN/dE$ in units of $\mathrm{GeV \, cm^{-2} \, s^{-1}}$ ($1 \, \mathrm{erg}\simeq624 \, \mathrm{GeV}$). This quantity measures the total energy flux in neutrinos, and it is useful as a performance indicator measuring the efficiency of neutrino production in the source. 

In order to see that, consider the transformation of the injection spectrum of the neutrinos $Q'_{\nu}$  (in units of $\mathrm{GeV^{-1} \, cm^{-3} \, s^{-1}}$) from a single source into a point source  flux  (in units of $\mathrm{GeV^{-1} \, cm^{-2} \, s^{-1}}$) at the detector $dN/dE$, which is (before flavor mixing and neglecting a possible Lorentz boost or beaming) given by (see, \eg, \Ref~\cite{Baerwald:2011ee})
\begin{equation}
 \frac{dN(E)}{dE} = V \,  \frac{(1+z)^2}{4 \pi d_L^2} \, Q'_{\nu} \, , \qquad E=\frac{1}{1+z} \, E' \, . \label{equ:boost}
\end{equation}
Here, $V$ is the  volume of the interaction region and $d_L(z)$ is the luminosity distance. For the energy flux density, one has
\begin{equation}
 \phi = \frac{L_\nu}{4 \pi d_L^2}  \, , \quad \text{where} \quad L_\nu=V \int E' Q'_\nu dE' \
\end{equation}
is the ``neutrino luminosity''. Since the neutrinos originate mostly from pion decays and take a certain fraction of the pion energy (about $1/4$ per produced neutrino for each charged pion), the neutrino luminosity is directly proportional to the (internal) luminosity of protons $L_{\mathrm{int}}$ (or the proton energy dissipated within a certain time frame $\Delta T$) and the fraction of the proton energy going into pion production, commonly denoted by $f_\pi$. This quantity is a measure of the efficiency of pion production.\footnote{It can be roughly approximated as $f_\pi \simeq 0.2  R/\lambda_{p\gamma}$ for the $\Delta$-resonance, where $\lambda_{p\gamma}=1/(n_\gamma \sigma)$ is the mean free path of the proton, $n_\gamma$ is the photon number density, and $\sigma$ is the interaction cross section. } Since a possibly emitted photon flux can be linked to $L_{\mathrm{int}}$  by energy equipartition arguments,  one has $\phi \propto f_\pi \times L_{\mathrm{int}} \propto f_\pi \times L_{\gamma}$, and $\phi$  is a measure for the pion production efficiency times luminosity of the source (if no photon counterpart is observed), or even the pion production efficiency itself (if a photon counterpart is observed). For example, for GRBs, this can be nicely seen in App.~A of \Ref~\cite{Abbasi:2009ig}, from which it is also clear that redshift and $\Gamma$ cancel if the neutrino flux is related to a photon observation (the neutrinos and photons will, to a first approximation, experience the same Lorentz boosts, beamings, and redshifts). Therefore, the sensitivity to $\phi$ for different spectra yielding the {\em same number of events} (as the fluxes do a the sensitivity limit) can be regarded as the prime performance indicator. Consider, for instance, two very similar sources producing neutrino spectra with slightly different shapes. Then the sensitivity to $\phi$, and therefore to the pion production efficiency, may be very different, and it is fair to say that one source can be better constrained than the other. Consider, on the other hand, two sources producing very different neutrino spectra. In that case, the sensitivity to $\phi$ could be even similar, which means a similar $f_\pi \times L_{\mathrm{int}}$ is required for detection. Our sensitivities to $\phi$ in the following sections have to be interpreted in that way.
Finally, note again that  our method does not predict the expected level of a neutrino flux, which means that in specific cases $f_\pi \times L_{\mathrm{int}}$ might actually be much higher than in other ones for astrophysical reasons. Here, however, here we discuss only the interplay between spectral shape and detector response.

\section{Constraints from IceCube}
\label{sec:constraints}

Here we present our main result, the sensitivity to the spectral shape of sources as a function of the Hillas parameters, and we discuss the dependence on the injection index $\alpha$, the impact of kaon decays, and $\nu_\tau$ detection via muon tracks.

\subsection{Sensitivity on Hillas plane}
\label{sec:hillassens}

\begin{figure}[tp]
\begin{center}
\includegraphics[width=0.7\textwidth]{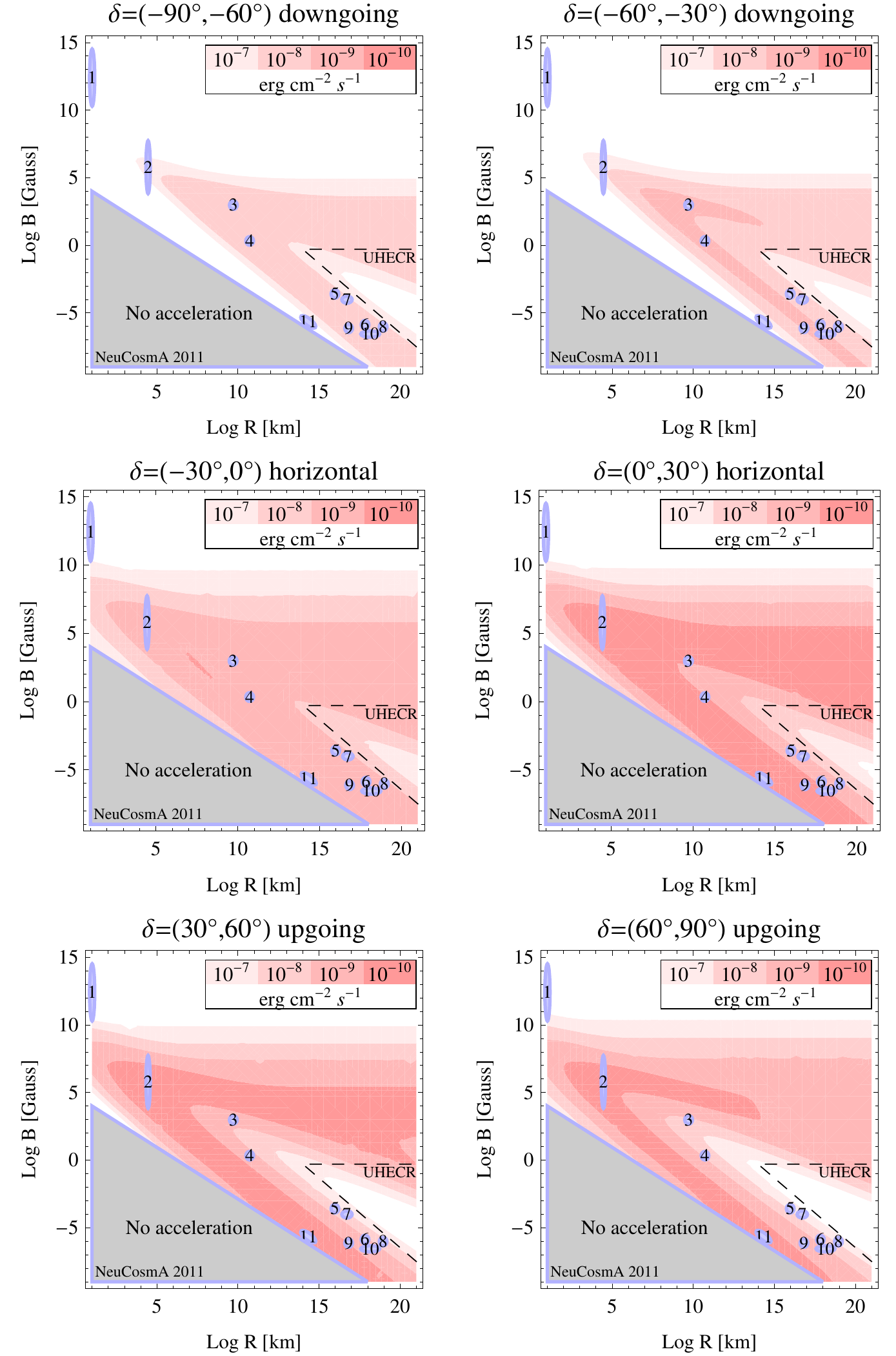}
\end{center}
\mycaption{\label{fig:hillassens} Energy flux density [$\mathrm{erg \, cm^{-2} \, s^{-1}}$] sensitivity as a function of $R$ and $B$ for different declination bands in IceCube-40 (90\% CL). The different contours (colors) correspond to the regions where a specific energy flux sensitivity can be exceeded. Here the injection index $\alpha=2$ is chosen. The dashed regions ``UHECR'' indicate where $10^{20} \, \mathrm{eV}$ cosmic ray protons are expected to be produced in the model. }
\end{figure}

For a comprehensive parameter space scan, we compute for each set $(R,B)$ in \figu{maxp} the $\nu_\mu$ spectrum including magnetic field and flavor effects. Then we normalize it with \equ{N} for each event type, as it is illustrated in \figu{spectra} for several examples, and we compute the energy flux density according to \equ{eflux}. The resulting sensitivity to $\phi$ is shown in \figu{hillassens} as a function of $R$ and $B$ for $\alpha=2$ and the different declination bands in the different panels.  The darkest regions mean highest sensitivities, as it is shown in the plot legends. For downgoing events, sensitivities as low as $10^{-8} \, \mathrm{erg \, cm^{-2} \, s^{-1}}$ can be achieved, for upgoing events, sensitivities as low as $10^{-10} \, \mathrm{erg \, cm^{-2} \, s^{-1}}$. The best parameter space coverage is actually obtained for near-horizontal upgoing events [$\delta=(0^\circ,30^\circ)$], which means that sources with relatively low $f_\pi \times L_{\mathrm{int}}$ can be detected there.

As one result, the detector responds very well to the usual suspects, such as AGN cores and jets (TP~3 and~4), and to sources on galactic scales, such as TP~9 and~10. For instance, in the upper right panel (downgoing events), these are just in the optimal region. However, the best absolute sensitivities are obtained for TP~2 and~11 rather than TP~3 and~4, see upgoing events in lower row. The reason is the relatively low optimal energy for upgoing events, see \figu{aeff}. In addition, TP~5,~7, and~8 are not within the optimal sensitivity ranges, since these spectra peak at relatively high energies (see, \eg, \figu{spectra} for TP~8). In fact, in our model, exactly these spectra stem from very high energy protons, \ie, these test points may be the best candidates to produce the highest-energetic cosmic ray protons (see regions marked ``UHECR'' in \figu{hillassens}). This just visualizes which might be an intrinsic feature of IceCube, which should be rather model-independent: the differential limits in \figu{aeff} peak at relatively low energies, while the neutrino energies for interactions of  $10^{21} \, \mathrm{eV}$ protons in the source may be expected at $10^{19}$ to $10^{20} \, \mathrm{eV}$ ($10^{10}$ to $10^{11} \, \mathrm{GeV}$). Therefore, IceCube may not be the best instrument to test the nature of the {\em highest-energetic} cosmic ray sources -- although it may compensate for that by its size. Note that this argument does not depend on the composition of the UHECR, since it depends on the energy of the particles, not on the region of the Hillas parameter space. Moreover, there is a part of the parameter space, which is very difficult to test: especially sources with $B > 10^5 \, \mathrm{Gauss}$ from downgoing events may be difficult to find in IceCube, which means that galactic sources with strong magnetic fields are {\em per se} difficult to access, unless they are very luminous.\footnote{Note that the energy losses of the secondaries decrease the fraction of energy going into neutrinos even further.} Here DeepCore may improve the sensitivity substantially, since it has a lower threshold.

\subsection{Dependence on injection index}
\label{sec:alpha}

\begin{figure}[t]
\begin{center}
\includegraphics[width=0.8\textwidth]{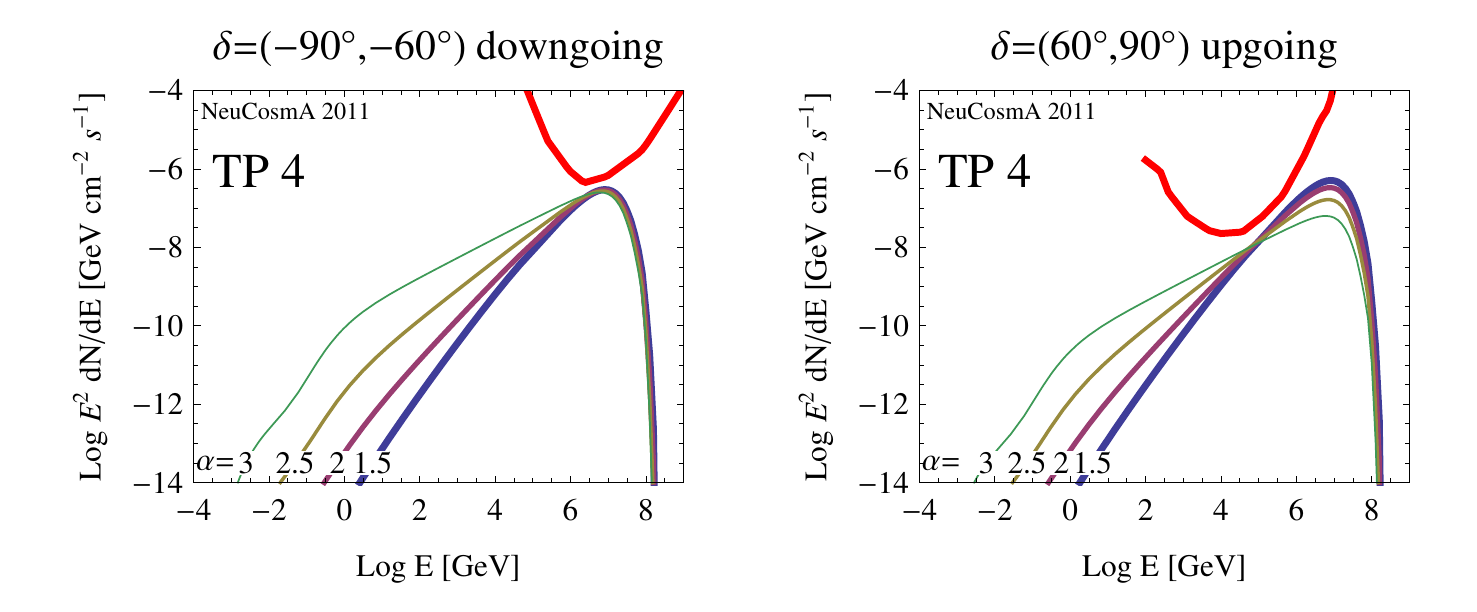}
\end{center}
\mycaption{\label{fig:alphaspec} Flux  limit for several different injection indices $\alpha=1.5$, $2$, $2.5$, and $3$ for test point~4 (AGN jets) and two different declination bands (90\% CL). The upper thick curves show the differential limits.}
\end{figure}

So far, we have chosen $\alpha=2$ as the universal injection index for protons and electrons, as it may be roughly expected from Fermi shock acceleration. This, however, is not the spectral index of the neutrino spectrum. In our model, the target photons are produced by synchrotron radiation from the co-accelerated electrons, and the (synchtrotron) energy losses of the electrons are taken into account. As a result, the spectral index of the neutrino spectrum is approximately $\alpha/2$ up to the maximal energy cutoff or the critical energy in \equ{ec}, where the secondary spectra become loss-steepended by synchrotron losses. This spectral shape is similar, for instance, to neutrinos from blazar jets, see, \eg, \Ref~\cite{Muecke:2002bi}. We show a corresponding example for 
TP~4 in \figu{alphaspec} for two different declination bands. In this figure, the sensitivities and differential limits are shown for several values of $\alpha$. One can easily see that the spectral index of the neutrino spectrum between about 1~GeV and 1~PeV is roughly $\alpha/2$. Since energy losses of the secondaries are small in this case, the spectral shape is relatively simple. For the case of the downgoing tracks (left panel), the peak of the spectrum coincides with the minimum of the differential limit. The energy flux sensitivity decreases as $\alpha$ increases, since more energy can be hidden in the neutrino spectrum at low energies.  For the upgoing events (right panel), the peak of the spectrum is found at higher energies than the minimum of the differential limit. In this case, increasing $\alpha$ means that the neutrino spectrum can be better constrained  and that the sensitivity, which is determined by the high energies, improves. 

\begin{figure}[t]
\begin{center}
\includegraphics[width=0.8\textwidth]{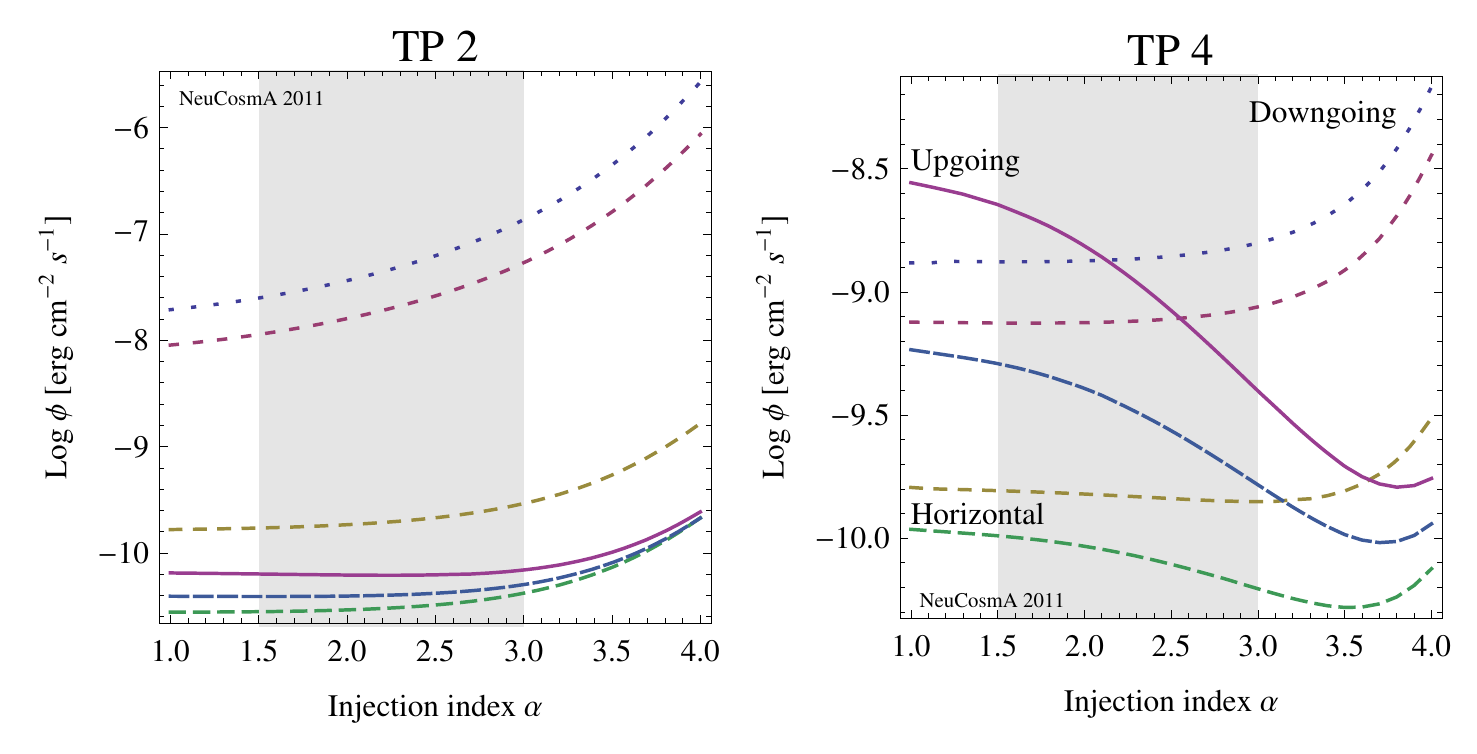}
\end{center}
\mycaption{\label{fig:alphadep} Energy flux density sensitivity as a function of the injection index $\alpha$ for two selected test points (90\% CL). The different curves represent the six different declination ranges used for IceCube-40, from downgoing [$\delta=(-90^\circ,-60^\circ)$] (dotted) to upgoing  [$\delta=(60^\circ,90^\circ)$] (solid) with decreasing dash gaps. The gray-shaded regions mark the $\alpha$-ranges which may be roughly plausible for Fermi shock acceleration.}
\end{figure}

It turns out that all test points in \figu{maxp} can be separated in two categories: TP~1,~2,~11, and~12 always follow the trend in \figu{alphaspec}, left panel, since the neutrino energies are relatively low. The other test points exhibit a behavior similar to TP~2: the optimum sensitivity depends in the declination of the source. We show the energy flux sensitivity as a function of the injection index $\alpha$ for two examples (TP~2 and TP~4) in \figu{alphadep}.  The different curves represent the six different declination ranges used for IceCube-40, from downgoing [$\delta=(-90^\circ,-60^\circ)$] (dotted) to upgoing [$\delta=(60^\circ,90^\circ)$] (solid) with decreasing dash gaps. In the right panel, we find the functional dependence for TP~4, which we have qualitatively described above for upgoing and downgoing events. Note that the actual dependence on $\alpha$ is relatively flat for the downgoing and quasi-horizontal events, at least for reasonable injection indices (gray-shaded range), whereas for upgoing events the sensitivity significantly improves with $\alpha$.\footnote{At about $\alpha=4$, a minimum is found in these cases. At this value, the neutrino spectral index is roughly two, which means that the slope in the $E^2 dN/dE$ plot changes at around this value from increasing to decreasing.} For TP~2 (left panel), low injection indices are always preferred, but the dependence on $\alpha$ is moderate. 

\subsection{Impact of neutrinos from neutron and kaon decays}
\label{sec:kaon}

\begin{figure}[t]
\begin{center}
\includegraphics[width=0.8\textwidth]{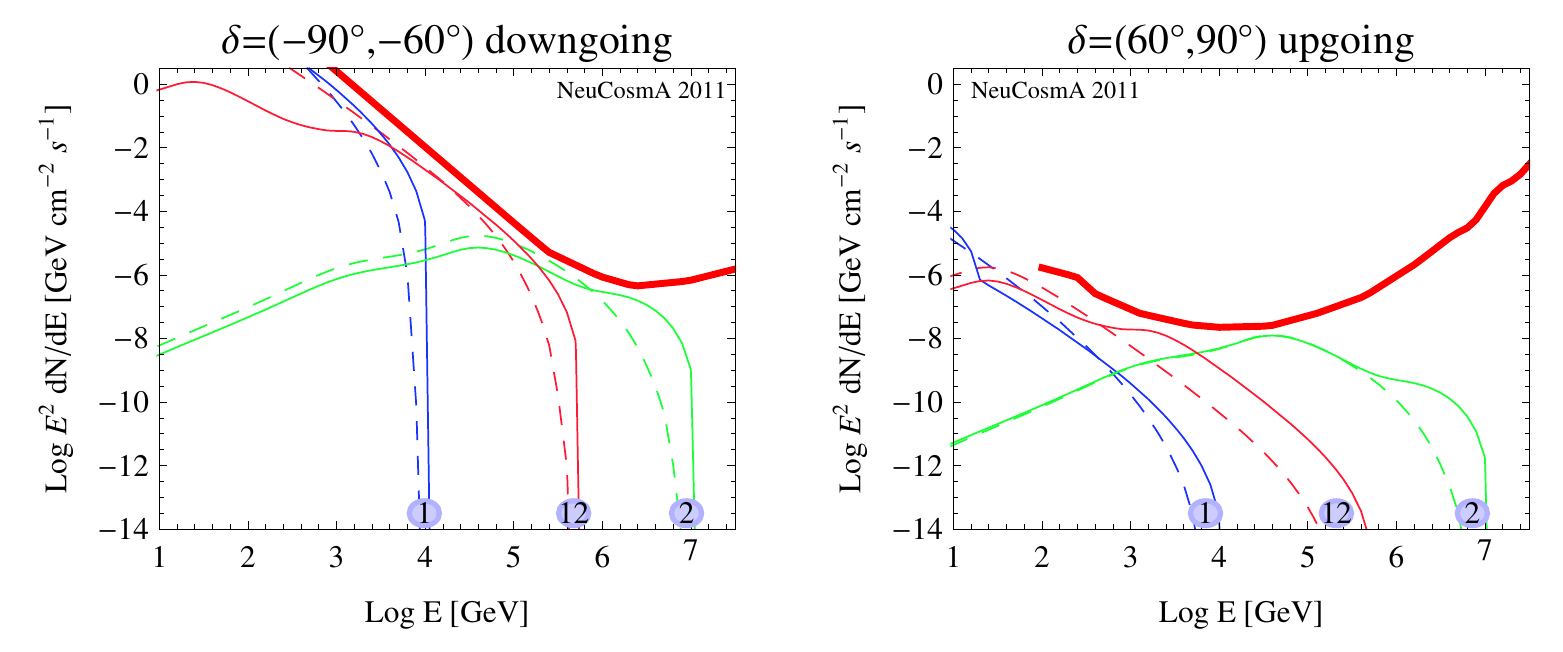}
\end{center}
\mycaption{\label{fig:impactnk} Impact of neutrinos from kaon decays: Limits for several $\nu_\mu$ spectra (including flavor mixing) for two different selected declination bands in IceCube-40 (90\% CL).  The solid curves include neutrinos from kaon (and neutron) decays, the dashed curves represent neutrinos from the pion decay chain only. The numbers of the individual curves correspond to the test points in \figu{maxp}. The thick curves show the  differential limits. Here the injection index $\alpha=2$ is chosen.}
\end{figure}

Here we discuss two neutrino fluxes often not taken into account: For sources optically thin to neutrons, most of the neutrons which are produced in photohadronic interactions (see \equ{photo1}) will decay into electron antineutrinos either inside or outside the source. The protons produced in these decays may contribute to the cosmic ray flux. The neutrinos carry only a very small fraction of the neutron energy, which means that this flux is normally present at very low energies. However, for sources with very strong magnetic fields, the neutrino flux from neutrons may actually be dominant, since the parents are electrically neutral. In addition, kaons can be produced by interactions such as \equ{photo2}. While the contribution of the neutrino flux from kaon decays is usually small, the higher value of the critical energy \equ{ec} compared to muons and pions may lead to a dominant neutrino flux from kaon decays at high energies for strong enough magnetic fields. 

All these effects are fully taken into consideration in our calculations. Nevertheless, we show in \figu{impactnk} the impact of these two neutrino fluxes at two examples. In this figure, 
the limits for several $\nu_\mu$ spectra  for two different selected declination bands  are shown.  The solid curves include neutrinos from kaon and neutron decays, the dashed curves show the limits if neutrinos from the pion decay chain only are considered. In the left panel, one can see a clear enhancement of the flux at high energies in all cases, which comes from the additional kaon decay component. For TP~2, the additional hump coincides with the differential limit minimum, which means that it contributes to the sensitivity (the solid curve is below the dashed curve for lower energies). The same applies to TP~1 and~12, although not so clearly visible. In the right panel, the contribution for TP~2 is negligible, since the solid and dashed curves coincide. There is, however, some contribution in the other two cases. For TP~1 one can also see the contribution from neutron decays dominating at low energies. However, while kaon decays help in parts of the parameter space, especially for downgoing events (where the differential minimum is at higher energies), the impact of neutron decays is, in general, small.

\subsection{Impact of muon tracks from $\boldsymbol{\nu_\tau}$?}
\label{sec:nutau}

\begin{figure}[t]
\begin{center}
\includegraphics[width=0.8\textwidth]{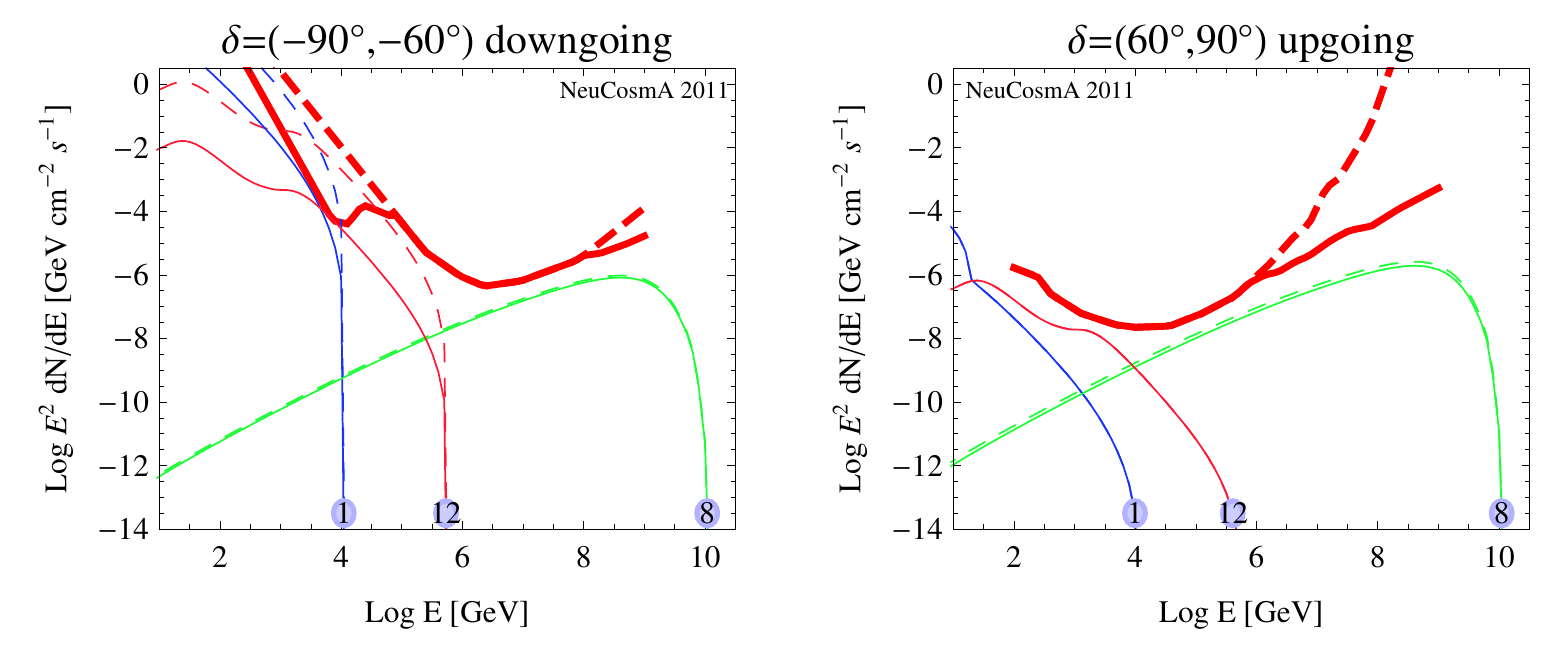}
\end{center}
\mycaption{\label{fig:impacttau} Impact of $\nu_\tau$ detection: Limits for several $\nu_\mu$ spectra (including flavor mixing) for two different selected declination bands in IceCube-40 (90\% CL).  The solid curves are based on the detection of muon tracks from both $\nu_\mu$ and $\nu_\tau$ assuming flavor equipartition, the dashed curves are based on the detection of muon tracks from $\nu_\mu$ only. The numbers of the individual curves correspond to the test points in \figu{maxp}. The thick curves show the  differential limits for both $\nu_\mu$ and $\nu_\tau$ (solid) and $\nu_\mu$ only (dashed). Here the injection index $\alpha=2$ is chosen.}
\end{figure}

In \figu{aeff}, we have compared the neutrino effective areas for muon tracks from $\nu_\mu$ and $\nu_\tau$. Assuming flavor equipartition between $\nu_\mu$ and $\nu_\tau$, are there parts of the parameter space where the muon tracks from $\nu_\tau$ actually limits the sensitivity? In fact, we show in \figu{impacttau} several examples where we find some impact. In this figure, the thick solid curves show the differential limits including the $\nu_\tau$ events (corresponding to the thick solid curves in \figu{aeff}), the thick dashed curves the contributions from $\nu_\mu$ only. The thin solid curves show selected spectra using the inclusive effective area, the thin dashed curves show the corresponding spectra for $\nu_\mu$ based events only. For the downgoing events (left panel), there is especially some impact at low energies, where the sensitivities can be significantly improved. For the upgoing events, the improved effective area for high energies has hardly any impact, at least for $\alpha=2$, for which the spectrum is parallel to the differential limit. Only for $\alpha < 2$, some improvement may be expected. Re-drawing \figu{hillassens} including the $\nu_\tau$ events, the result mostly deviates a bit in the high energy region in the lower right. The effect at low energies is hardly visible, because the absolute sensitivities for TP~1 and~12 are worse than the shown contours. 

\section{Complementarity to other experiments}
\label{sec:compl}

Here we qualitatively point out the complementarity among different data and different experiments, and we comment on the potential of future experiments. There are a number of high-energy neutrino data, such as from AMANDA~\cite{Ahrens:2003pv,Ackermann:2004aga},  IceCube-22 cascades~\cite{Abbasi:2011ui}, Auger~\cite{Abraham:2009uy}, RICE~\cite{Kravchenko:2006qc}, ANITA~\cite{Gorham:2010kv}, to name a few examples. Most of these data (except from AMANDA) have been applied to diffuse flux limits, which makes a comparison to the IceCube point source results, which are discussed in this study, {\em per se} difficult. While AMANDA has, in principle, a lower threshold than IceCube, it is not clear from the present literature if the neutrino effective areas are significantly better at low energies than for IceCube-40, since the larger volume may compensate for that. The RICE and ANITA experiments are based on radio detection initiated by cascades in the Antarctic ice. In these cases, as for cascades in IceCube, it is difficult to quantify  the performance for a particular flavor. For example, in \Ref~\cite{Abbasi:2011ui} (IceCube cascades), a flavor composition of 40\% electron neutrinos, 45\% tau neutrinos, and 15\% muon neutrinos was given for an $E^{-2}$ extragalactic test flux; see also discussion in \App~I in \Ref~\cite{Kravchenko:2006qc} (RICE). The only exception in this list is Auger: Earth-skimming $\nu_\tau$ may produce tau leptons in the Earth, which may escape the Earth after energy losses, decay in the atmosphere, and produce an extensive quasi-horizontal, slightly upgoing air shower detectable by the Auger surface detector. Since this signal is practically flavor-clean, we illustrate our main points with 2004--2008 data from the Auger experiment~\cite{Abraham:2009uy}, which appears to be simplest example. However, note that also the other experiments imply complementary information. 

\subsection{Earth-skimming neutrinos in Auger}
\label{sec:auger}

\begin{figure}[t]
\begin{center}
\includegraphics[width=0.4\textwidth]{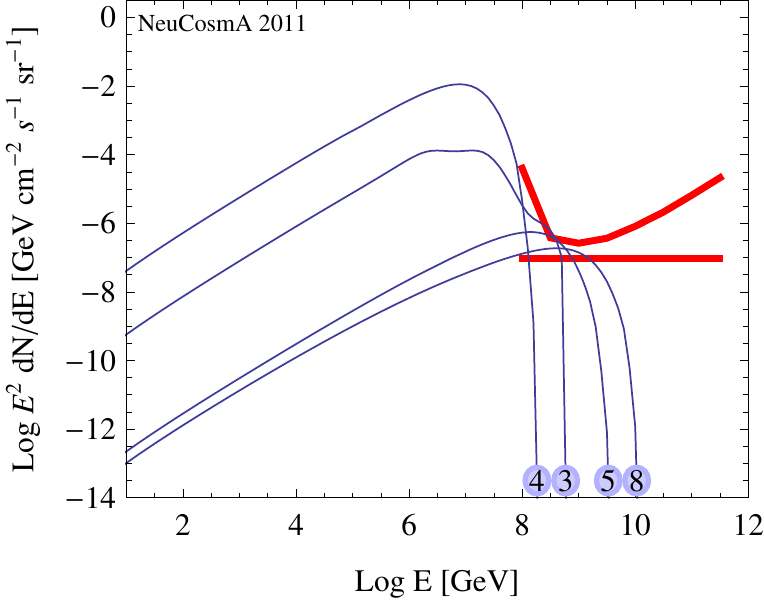}
\end{center}
\mycaption{\label{fig:spectraauger} Limits for selected $\nu_\tau$ spectra (including flavor mixing) for Auger 2004--2008 data (90\% CL). The numbers of the individual curves correspond to the test points in \figu{maxp}. The thick line show the limit for an $E^{-2}$ flux (in the dominant energy range), and the thick curve the differential limit. Here the injection index $\alpha=2$ is chosen. }
\end{figure}

The sensitivity for arbitrary fluxes for Auger can be very easily obtained from \equ{N}. We use the ``conservative systematics'' exposure from table~III of \Ref~\cite{Abraham:2009uy} for our simulation. As the main differences to IceCube, diffuse flux limits are discussed, and the event sample is free of backgrounds from the atmospheric neutrino flux in that energy range. Since the viewing window constantly changes, there is a strong declination dependence of the sensitivity. However, for the sake of comparison, one may compute a ``quasi-point source'' flux by multiplying with the diffuse flux with the viewing window solid angle $d \Omega \sim 0.6$.\footnote{The solid angle $d \Omega \sim 2 \pi \int d \cos \theta \sim 0.6$, since $\theta$ is integrated from $\pi/2$ to $\pi/2+\alpha_m$ with $\alpha_m=0.1$; see discussion after Eq.~4 in \Ref~\cite{Abraham:2009uy}.} Since this number is of order unity, we use the diffuse flux in the case of Auger directly. However, one should keep in mind this qualitative difference when one compares the numbers, which cannot be easily avoided.\footnote{A point source analysis at Auger is presently being performed~\cite{Guardincerri:2011pf}, using 2004--2010 data. For a more detailed analysis than the one presented here, including the strong declination dependence,  the exposure as a function of energy and declination would be needed, which is currently not publically available. However, from the comparison with \Ref~\cite{Guardincerri:2011pf}, our sensitivity with the 2004--2008 data roughly corresponds to a point source sensitivity  at the declination $50^\circ \lesssim | \delta | \lesssim 55^\circ$ with the updated 2004--2010 data set (replace units by $\mathrm{GeV} \, \mathrm{cm}^{-2} \, \mathrm{s}^{-1}$ on the vertical axis in \figu{spectraauger}).} 

We show in \figu{spectraauger} the limits for selected $\nu_\tau$ spectra (including flavor mixing) for Auger 2004--2008 data (90\% CL). The numbers of the individual curves correspond to the test points in \figu{maxp}. The thick lines show the limit for an $E^{-2}$ flux (in the dominant energy range), and the thick curves the differential limit. Comparing to \figu{spectra}, one can easily see that especially the spectra peaking at very high energies can be very well constrained. Take, for instance, TP~8, and compare the result to \figu{spectra}, lower left panel. It is obvious that the normalization can in this case be better constrained by Auger, and, consequently, the energy flux density.  We have also tested the impact of neutrinos from kaon decays here. Because the neutrinos from kaon decays show up at high energies and the Auger sensitivity (see differential limit) is dominant at about $10^9 \, \mathrm{GeV}$, it turns out that the kaon component can be especially important in that case, see also \Ref~\cite{Asano:2006zzb}. One example, TP~3, is shown in \figu{spectraauger}, where the rightmost hump is the additional contribution from kaon decays which clearly limits the sensitivity.

\subsection{Complementarity among different data and experiments}
\label{sec:compls}

\begin{figure}[t]
\begin{center}
\includegraphics[width=0.47\textwidth]{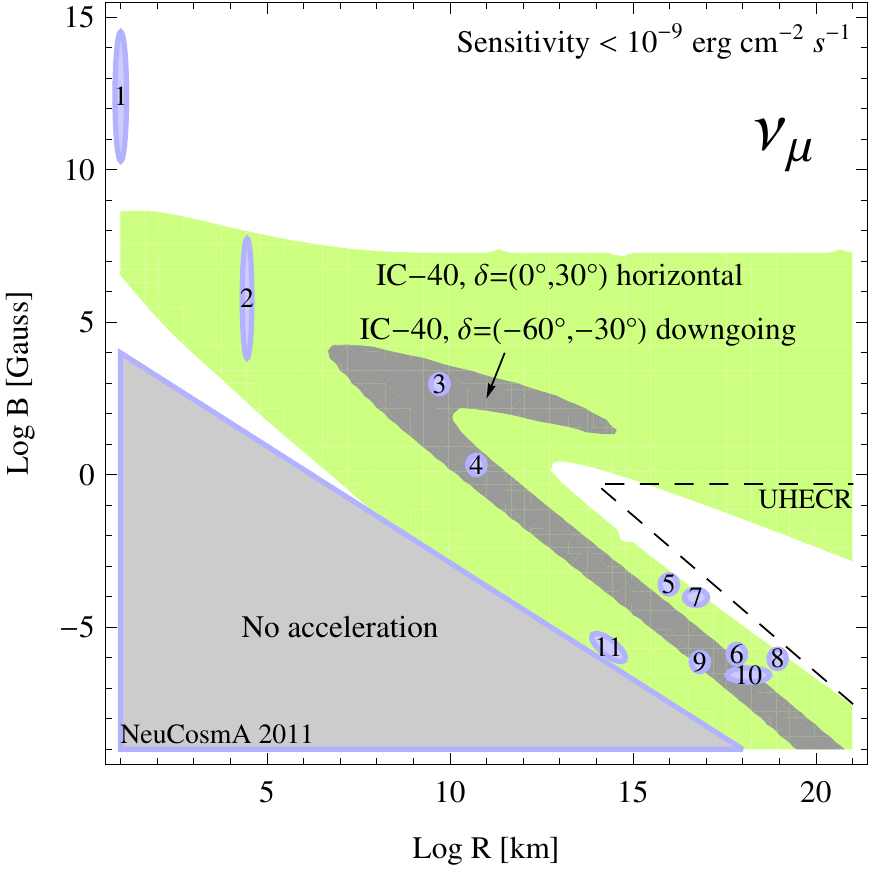} \hspace*{0.03\textwidth}
\includegraphics[width=0.47\textwidth]{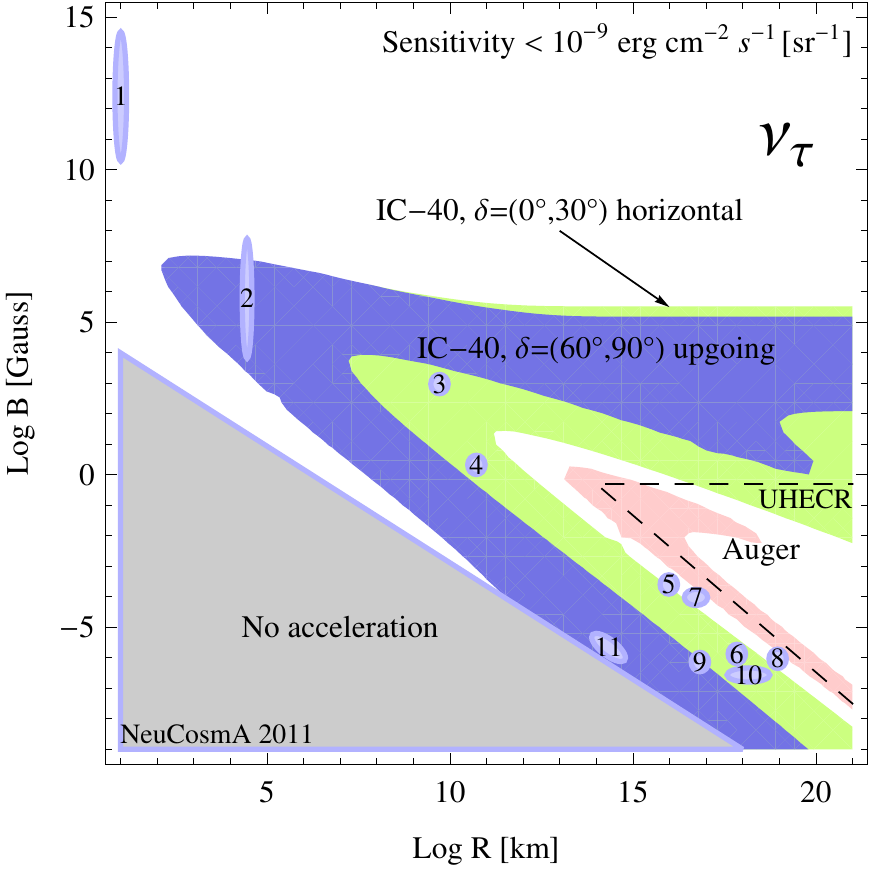}
\end{center}
\mycaption{\label{fig:comp} Regions where the sensitivity exceeds $10^{-9} \, \mathrm{erg \, cm^{-2} \, s^{-1} \, [sr^{-1}]}$ for $\nu_\mu$ detection (left panel) and $\nu_\tau$ detection (right panel) for several selected data samples (90\% CL, $\alpha=2$). The dashed regions ``UHECR'' indicate where $10^{20} \, \mathrm{eV}$ cosmic ray protons are expected to be produced in the model. See main text for details.}
\end{figure}

In order to illustrate the complementarity of different data and experiments, we show in \figu{comp} the regions where the sensitivity exceeds $10^{-9} \, \mathrm{erg \, cm^{-2} \, s^{-1} \, [sr^{-1}]}$ for $\nu_\mu$ detection (left panel) and $\nu_\tau$ detection (right panel) for several selected data samples. In the case of IceCube, only muon tracks from $\nu_\mu$ or $\nu_\tau$ have been used, see neutrino effective areas in \figu{aeff}. 
In the case of Auger, the diffuse flux limit for Earth-skimming tau neutrinos has been used.  
Note that (at least if the $\tau$ track cannot be separated) IceCube cannot distinguish the original flavor of the neutrino leading to a muon track, while Auger can.
With this figure, we can qualitatively illustrate a number of points:

\subsubsection*{Viewing window complementarity}

Different data samples correspond to different viewing angles. In the case of IceCube, the upgoing events test the northern sky, the downgoing events the southern sky. From the left panel, we clearly see the better sensitivity of, for instance, the quasi-horizontal upgoing events compared to the downgoing ones. For almost vertical downgoing events, the sensitivity is even poorer (\cf, \figu{hillassens}). New experiments built in the Northern hemisphere will complement this sensitivity, such as ANTARES or KM3NeT. 

\subsubsection*{Energy range complementarity}

Different experiments or data will test different energy ranges. For example, from the right panel, IceCube can cover a wide region of the parameter space. However, for the high $B$ (low energy) region, which is almost empty, experiments with a lower threshold will be needed. Especially the DeepCore part of IceCube is expected to extend into this region. In the lower right corner marked ``UHECR'', however, where the highest-energetic cosmic ray sources are suspected for protons, in fact Auger provides the best sensitivity, given our assumptions. Therefore, Auger and the radio-detection experiments may in fact be promising techniques to reveal the nature of these. Especially the Auger North project~\cite{Blumer:2010zza} may achieve a substantially higher exposure in exactly that energy range~\cite{Kotera:2010yn}, as well as the JEM-EUSO project~\cite{MedinaTanco:2009tp}.  

\subsubsection*{Flavor complementarity}

In \figu{comp}, the left and right panel correspond to different neutrino flavors at the detector. If, for some reason, the equipartition between $\nu_\mu$ and $\nu_\tau$ is severely perturbed, it is first of all interesting that IceCube can, at least for upgoing or quasi-horizontal upgoing events, still test most of the parameter space relatively well (right panel). Of course, the Auger limit applies to $\nu_\tau$ only, but in the case of flavor equipartition it can be directly applied to $\nu_\mu$.  It is now an interesting theoretical question when one may expect a strong deviation from the equipartition between $\nu_\mu$ and $\nu_\tau$, which is a consequence of nearly maximal atmospheric mixing, at the detector. 

In the Standard Model including massive neutrinos, flavor equipartition between these two fluxes relies on two necessary conditions: $\theta_{13}=0$ and $\theta_{23}=\pi/4$, which guarantee flavor equipartition regardless of the initial flavor composition at the source. Note that exact tri-bimaximal mixings ($\sin^2 \theta_{12}=1/3$) are not necessary. 
We  have chosen these values in this study, which means that all of our results are exactly $\nu_\mu$-$\nu_\tau$ symmetric. However, there may be deviations from these values.
Therefore, we have checked that for arbitrary initial flavor compositions (without $\nu_\tau$ contamination at the source) the ratio between $\nu_\mu$ and $\nu_\tau$ can, at most, vary between about 0.5 and 2 for the currently allowed $3 \sigma$ ranges of the mixing parameters~\cite{Schwetz:2008er}. This means that within the Standard Model, the $\nu_\mu$ and $\nu_\tau$ fluxes at the detector have to be equal up to a factor of two, and any limit on $\nu_\mu$ ($\nu_\tau$) can be directly translated into a limit for $\nu_\tau$ ($\nu_\mu$) within a factor of two.\footnote{The deviations from flavor equipartition between $\nu_\mu$ and $\nu_\tau$ can in fact be largest if the initial flavor composition is dominated by $\nu_e$.} On logarithmic scales, flavor equipartition is therefore guaranteed.

This picture may change if physics beyond the Standard Model (BSM) is considered. First of all, 
strong perturbations of the flavor equipartition between $\nu_\mu$ and $\nu_\tau$ at the detector may be unlikely if the physics BSM causes effects at the production or propagation of the neutrinos, since maximal atmospheric mixing will lead to equilibration of the two flavors again. Nevertheless, such exotic scenarios have been discussed in the literature, such as in the context CPT violation, see \Ref~\cite{Barenboim:2003jm} for a discussion. Any new physics interaction at the detection (or, for upgoing events, in the Earth) may be most plausible reason to perturb this ratio. Examples are non-standard interactions at the detection,  such as $\epsilon^{ud}_{\mu \tau}$, which are, however,  limited to about 10\% compared to the Standard Model~\cite{Biggio:2009nt}. Another possibility is a possible superluminal motion of muon neutrinos, OPERA has recently claimed at the $6\sigma$ confidence level~\cite{Adam:2011zb}, if a flavor dependent effect is at work which advances or delays one flavor compared to the others (see also \Ref~\cite{Autiero:2011hh}).
In addition, it is not clear how the deep inelastic scattering cross sections evolve to high energies. Therefore, considering different flavor may still be considered complementary.

\section{Summary and conclusions}
\label{sec:summary}

The main motivation of this study has been the discussion of the interplay between spectral shape and detector response at neutrino telescopes beyond a simple $E^{-2}$ assumption for the neutrino flux, from a particle physics perspective. Several effects often not taken into account have been included, such as magnetic field effects on the secondaries and flavor mixing. Especially, the impact of neutrinos from neutron and kaon decays has been discussed, and the impact of $\tau$ decays into muon tracks in the detector. As data samples, we have used 2008--2009 data from IceCube-40 for time-integrated point source searches, and 2004--2008 data from Auger. As parameter space of interest, we have used the Hillas plot, described by the parameters $R$ (size of the acceleration region) and $B$ (magnetic field). For the description of the neutrino spectra, we have used a self-consistent model for the neutrino production, in which protons interact with synchrotron photons from co-accelerated electrons (positrons)~\cite{Hummer:2010ai}. While this model certainly does not apply to all spectra on the Hillas plot, it has been useful to illustrate some of the main qualitative points as a function of $R$ and $B$.

In order to compare the detector response to different spectral shapes, we have used the sensitivity to the energy flux density. The energy flux density has been introduced as a measure for the source luminosity $L$ times pion production efficiency $f_\pi$ (if no photon counterpart is observed), or even $f_\pi$ directly (if the neutrino spectrum is compared to a photon observation). Here the pion production efficiency is the fraction of proton energy dumped into pion production. This means that the ``sensitivity'' in this work can be roughly interpreted as the sensitivity to $f_\pi$. Conversely, a source in a region of high sensitivity will be easier detectable, \ie, for lower luminosities and $f_\pi$, than a source in a region of low sensitivity. Note that we do not make any prediction for the expected level of the neutrino flux, as it can only come from astrophysical arguments.

Since a particular source will be seen with a specific declination in IceCube, the final (shape) sensitivity will be declination-dependent.  As one result, we have confirmed that IceCube has excellent sensitivity to spectral shapes corresponding to AGN cores or jets for the model considered, which are among the usual suspects for significant neutrino production, for all declinations. For example, they are found to be within the optimal region for downgoing tracks in IceCube. However, even better absolute sensitivities are obtained for quasi-horizonal and upgoing events for the spectral shapes from potential source classes such as white dwarfs, whereas the AGN blazar spectra may peak at a bit too high energies for these source declinations.
The strong magnetic field region, for which the maximal proton energy is synchrotron limited, cannot be easily tested by IceCube because of the relatively high threshold. Therefore, galactic sources with high magnetic fields and significant neutrino production are {\em per se} difficult to find. Data from DeepCore may significantly improve this parameter region. Note that some optimization of the detector response may be performed for specific spectral shapes, which we have not taken into account.

 The parameter space region from which the highest-energetic cosmic ray protons may be expected, \ie, $E \sim 10^{20} \, \mathrm{eV}$, is in fact somewhat better covered by Auger, at least in principle. This can be understood from generic arguments, recovered in the model used: $10^{20}  \, \mathrm{eV}$ protons colliding with much less energetic target photons will lead to neutrinos of about $10^{9}  \, \mathrm{GeV}$ to $10^{10}  \, \mathrm{GeV}$, which is just around the optimum of the differential limit of Auger. The conclusion holds for neutrino spectra significantly harder than $E^{-2}$, which is a common expectation of the target photons come from synchrotron emission. Therefore, the common picture of an $E^{-2}$ neutrino flux can be clearly misleading, and the search for neutrinos from the sources of the highest-energetic cosmic rays may greatly benefit from future experiments, such as Auger North or JEM-EUSO. This argument does not depend on the composition of the highest-energetic cosmic rays, since it is related to their energy, not their nature.

We have also tested the dependence on other model parameters, such as the injection index $\alpha$, which has some impact, especially for  spectra extending to high energies for upgoing events in IceCube. Again, it is clear that general conclusions on astrophysical sources cannot be drawn independent of the neutrino spectral shape because of the interplay with the detector response. We have, for instance, shown several examples where the neutrino spectrum peaks off the energy range IceCube is most sensitive to, which means that rather large $f_\pi  \times L$ are required in order to be detectable. In turn, this means that a source may not be found in spite of a correct astrophysical prediction of the flux, often based on energy equipartition arguments, simply because the flux shape does not match the detector response.

As far as the particle physics is concerned, we have shown that neutrinos from kaon decays improve the sensitivities in cases where magnetic fields lead to an additional kaon decay hump in the spectrum and this hump coincides with the optimal sensitivity range. This effect can be especially prominent in Auger, since the kaon decay part shows up at the high energy end of the spectrum. In IceCube, the contribution from $\nu_\tau$-induced muon tracks (via the leptonic $\tau$ decay channel) improves the sensitivities in parts of the parameter space if flavor equipartition between $\nu_\mu$ and $\nu_\tau$ at the detector is assumed, which means that it should be not {\em a priori} neglected.

Finally, we have emphasized the complementarity among different event samples and experiments, such as with respect to viewing window, accessible energy ranges, and different measured flavors. For example, we have demonstrated that even if the equipartition between $\nu_\mu$ and $\nu_\tau$ is strongly perturbed,  IceCube can cover most of the discussed parameter space at least for upgoing events already from the $\nu_\tau$-induced contribution to the muon tracks. While such a perturbation can only be expected within the (neutrino) Standard Model up to a factor of two for the current mixing parameter uncertainties, new physics effects may be in charge of larger deviations. 

We conclude that the interplay between spectral shape and detector response is important for the detection of astrophysical neutrino sources. While the common assumption of an $E^{-2}$ neutrino flux is well motivated for GRBs, it is well known not to apply to different source classes, such as AGNs. Since the detector response strongly depends on the shape of the neutrino spectrum, the amount of energy which is needed to be dumped into pion production at the source to guarantee a neutrino detection may have to be actually higher than assumed in many astrophysical source models.

\subsubsection*{Acknowledgments}

I would like to thank Philipp Baerwald, Sandhya Choubey, Raj Gandhi, Svenja H{\"u}mmer, Michele Maltoni, and Carlos Yaguna for useful discussions and comments.
This work has been supported by DFG grants WI 2639/2-1 and WI 2639/3-1.


\end{document}